\documentclass[nohyper,letterpaper]{JHEP3}

\usepackage{epsfig,multicol,bbm}
\usepackage{mathrsfs,amsmath,amssymb,graphicx}

\newcommand\fverb{\setbox\fverbbox=\hbox\bgroup\verb}
\newcommand\fverbdo{\egroup\medskip\noindent\fbox{\unhbox\fverbbox}\ }
\newcommand\fverbit{\egroup\item[\fbox{\unhbox\fverbbox}]}
\newbox\fverbbox

\newcommand{\de}{\mathrm d}
\newcommand{\oi}{\sum_i\Omega_i}
\newcommand{\om}{{\Omega_m}}

\newcommand{\orad}{{\Omega_r}}
\newcommand{\ok}{{\Omega_K}}

\newcommand{\os}{{\Omega_\sigma}}

\newcommand{\ho}{{H_0}}
\newcommand{\rc}{{r_c}}

\title{Testing a Phenomenologically Extended DGP Model with Upcoming Weak Lensing Surveys}

\author{Stefano Camera,$^{a,b}$ Antonaldo Diaferio$^{a,b,c}$ and Vincenzo F. Cardone$^{d,e}$\\$^a$Dipartimento di Fisica Generale ``A. Avogadro'', Universit\`a di Torino, Torino, Italy\\$^b$Istituto Nazionale di Fisica Nucleare (INFN), Sezione di Torino, Torino, Italy\\$^c$Harvard-Smithsonian Center for Astrophysics, Cambridge, MA, USA\\$^d$Dipartimento di Scienze e Tecnologie per l'Ambiente e il Territorio, Universit\`a degli Studi del Molise, Isernia, Italy\\$^e$Dipartimento di Scienze Fisiche, Universit\`a di Napoli, Napoli, Italy\\E-mail: \email{camera@ph.unito.it, diaferio@ph.unito.it, winnyenodrac@gmail.com}}

\preprint{\today}

\abstract{A phenomenological extension of the well-known brane-world cosmology of Dvali, Gabadadze and Porrati (eDGP) has recently been proposed. In this model, a cosmological-constant-like term is explicitly present as a non-vanishing tension $\sigma$ on the brane, and an extra parameter $\alpha$ tunes the cross-over scale $\rc$, the scale at which higher dimensional gravity effects become non negligible. Since the Hubble parameter in this cosmology reproduces the same $\Lambda$CDM expansion history, we study how upcoming weak lensing surveys, such as Euclid and DES (Dark Energy Survey), can confirm or rule out this class of models. We perform Markov Chain Monte Carlo simulations to determine the parameters of the model, using Type Ia Supernov\ae, $H(z)$ data, Gamma Ray Bursts and Baryon Acoustic Oscillations. We also fit the power spectrum of the temperature anisotropies of the Cosmic Microwave Background to obtain the correct normalisation for the density perturbation power spectrum. Then, we compute the matter and the cosmic shear power spectra, both in the linear and non-linear r\'egimes. The latter is calculated with the two different approaches of Hu and Sawicki (2007) (HS) and Khoury and Wyman (2009) (KW). With the eDGP parameters coming from the Markov Chains, KW reproduces the $\Lambda$CDM matter power spectrum at both linear and non-linear scales and the $\Lambda$CDM and eDGP shear signals are degenerate. This result does not hold with the HS prescription: Euclid can distinguish the eDGP model from $\Lambda$CDM because their expected power spectra roughly differ by the $3\sigma$ uncertainty in the angular scale range $700\lesssim\ell\lesssim3000$; on the contrary, the two models differ at most by the $1\sigma$ uncertainty over the range $500\lesssim\ell\lesssim3000$ in the DES experiment and they are virtually indistinguishable.}


\begin{document}

\section{Introduction}
The current $\Lambda$CDM concordance cosmological model excellently reproduces a large number of cosmological datasets. However, in order to explain the dynamics of galaxies and galaxy clusters and the Large-Scale Structures (LSS) of the Universe \cite{Zwicky:1933gu,Zwicky:1937zza,Dodelson:2001ux,Hawkins:2002sg,Spergel:2006hy,Riess:2006fw}, the observed pattern of the temperature anisotropies of the Cosmic Microwave Background (CMB) \cite{Hinshaw:2008kr,Komatsu:2008hk} and the present accelerated expansion of the cosmos \cite{Riess:1998cb,Knop:2003iy,Riess:2004n,Riess:2006fw,Larson:2010gs}, we need to postulate that more than $80\%$ of the matter in the Universe is made of weakly-interacting non-baryonic Dark Matter (DM) and that $\sim70\%$ of the total energy budget is due to a particular negative-pressure Dark Energy (DE) in the form of a cosmological constant $\Lambda$.

However, if one interprets the cosmological constant as the energy density of the vacuum of a quantum field, its measured value is more than hundred orders of magnitude smaller than what is predicted by quantum field theory (for a different perspective, see\cite{Bianchi:2010uw}). To avoid this problem, a number of alternative cosmological models has arisen over the last decades. They mainly try to reproduce the current data without adding any DM and/or DE, but rather by modifying or generalizing the law of gravity.

Here, we focus on a particular class of these theories, known as brane-world cosmologies. They consider the Universe as a differential manifold, called bulk, with more than four dimensions, which contains a 4D submanifold, the brane, with three spatial and one temporal dimensions; this submanifold is our spacetime. The model proposed by Dvali, Gabadadze and Porrati \cite{Dvali:2000rv,Dvali:2000hr} is of particular interest, as its 5D Minkowskian bulk is able to induce a late-time acceleration of the Universe expansion on the brane in agreement with current observations. However, the growth of cosmological structures in the DGP model is strongly suppressed compared to the $\Lambda$CDM model \cite{Lue:2002sw,Tanaka:2003zb,Lue:2004rj,Koyama:2005kd,Cardoso:2007xc}, and this also reflects on the DGP weak lensing power spectrum \cite{Ishak:2005zs,Thomas:2008tp}.

A phenomenological extension of the DGP model has recently been proposed \cite{Dvali:2003rk,Dvali:2006su,Schaefer:2007nf,Afshordi:2008rd,Kobayashi:2009da}. In this extension, the DGP cross-over length $\rc$, that determines the scale at which higher-dimensional gravitational effects become important, is tuned by a new parameter $\alpha$, and a cosmological-constant term is explicitly present as a non-vanishing tension $\sigma$ on the brane. These models are free of ghost-like instabilities and one can obtain such a model with the so-called cascading gravity framework \cite{deRham:2007xp,deRham:2007rw,deRham:2008qx}. In this construction, our 4D brane lies within a succession of higher-dimensional DGP branes, embedded in each other within a flat bulk spacetime.

The eDGP model we use has been proposed as a generalisation of the DGP model in a spatially flat Universe, but it might be also worth investigating the non-flat case. For the standard DGP, which presents strong tension between SNeIa data and CMB distance indicators, open models provide a better fit to the geometric data, because the distance to the last scattering surface is not constrained by the flat geometry. By introducing $\ok$ as an additional free parameter, the constraints on the cosmological parameters would be weaker and, consequently, the importance of the tests on the growth of structures would increase. Unfortunately,  DGP generalisations for the non-flat case similar or related to the model we examine here still lack.

In the present work, we explore this extended DGP (eDGP) model in detail. We perform Monte Carlo Markov Chains to obtain the best-fit value of the cosmological parameters. To this end, we use a number of observational datasets, which have been shown to be useful in the case of modified gravity models \cite{Cardone:2009yt}. Specifically, we use: the Hubble constant $\ho$, estimated by the HST Project using a well calibrated set of local distances \cite{Freedman:2000cf}; the latest 397 Type Ia Supernova (SNIa) distance moduli of the Constitution sample, compiled in Ref.~\cite{Hicken:2009dk}; the updated Gamma Ray Burst (GRB) Hubble diagram recently presented in Ref.~\cite{Cardone:2009mr}; Baryon Acoustic Oscillation (BAO) data from the seventh release of the Sloan Digital Sky Survey (SDSS); and distance priors from WMAP5. We also run the chains for the $\Lambda$CDM model, to compare our results. We also perform a chi-squared analysis on the temperature anisotropy power spectrum of the CMB with the most recent data coming from WMAP7 \cite{Larson:2010gs}, to estimate the normalisation of the power spectrum of the matter density perturbations. 

We calculate the matter power spectrum in both the linear and the non-linear r\'egime of perturbations, the latter with both the recent fitting formul\ae\ of Khoury and Wyman \cite{Khoury:2009tk}, who obtained a useful modification of the \textsc{halofit} procedure \cite{Smith:2002dz} for the eDGP model, and the interpolation proposed by Hu and Sawicki~\cite{Hu:2007pj}.

Finally, we compute the weak lensing convergence power spectrum. Weak lensing is a powerful tool \cite{Hu:2000ee,Hu:2001fb,Hu:1998az,Munshi:2006fn,Schmidt:2008hc}, and it is particularly useful to aim of testing and constraining modified/alternative cosmological models \cite{Jain:2007yk,Tsujikawa:2008in,Thomas:2008tp,Camera:2009uz,Shapiro:2010si,Camera:2010wm,Martinelli:2010wn}. Gravitational lens effects are due to the deflection of light occurring when photons travel near matter, i.e. in the presence of a non-negligible gravitational field. Therefore, the cosmic convergence and shear encapsulate information about both the sources that emit the light and the structures that the photons cross before arriving at the telescope: weak lensing enables us to explore both the basis of the cosmological model and the LSS of the Universe. In other words, it provides information about geometry and dynamics. The study of the power spectrum of weak lensing can be a crucial test, particularly with a view to upcoming weak lensing surveys such as the Dark Energy Survey (DES)\footnote{http://www.darkenergysurvey.org} \cite{Abbott:2005bi,2010AAS...21547704K}, the Panoramic Survey Telescope Rapid Response System (Pan-STARRS)\footnote{http://pan-starrs.ifa.hawaii.edu} \cite{2002SPIE.4836..154K,2002AAS...20112207K} and future space surveys such as Euclid\footnote{http://sci.esa.int/science-e/www/area/index.cfm?fareaid=102} \cite{2008SPIE.7010E..38R,Refregier:2010ss}.

The paper is organized as follows: in \S~\ref{edgp} the eDGP model is outlined; in \S~\ref{estimation} we present the cosmological parameters obtained with Monte Carlo Markov Chains and the normalisation of the power spectrum of the density perturbations derived from the fit to the  temperature anisotropies of the CMB; in \S~\ref{observables} we show our results: the matter power spectrum (\S~\ref{matterpower}) and the weak lensing shear (\S~\ref{weaklensing}). Finally, in \S~\ref{conclusions} conclusions are drawn.

\section{The extended DGP model}\label{edgp}
In the DGP model \cite{Dvali:2000rv,Dvali:2000hr,Deffayet:2000uy,Lue:2004rj}, the bulk has five dimensions and is Minkowskian, and the brane is flat. In this case, the more general homogeneous and isotropic metric has three different scale factors, one for the time $t$, one for the three spatial coordinates $\mathbf x=(r,\vartheta,\varphi)$ and one for the fifth dimension $y$, and takes the form\footnote{We use units such that $c=1$ and uppercase Latin indices run over bulk dimensions, Greek indices over 4D spacetime dimensions, whereas lowercase Latin indices label spatial coordinates.}
\begin{equation}
g_{AB}\,\de x^A\,\de x^B=-N^2(r,y)\,\de t^2+A^2(r,y)\,\de r^2+B^2(r,y)\,\de\Omega^2+\de y^2,\label{sovradensita}
\end{equation}
where $\de\Omega^2=r^2\,\de\vartheta^2+r^2\sin^2\vartheta\,\de\varphi^2$ is the differential solid angle.

In this model, gravity is general relativity (GR), but the Hilbert-Einstein Lagrangian is given in five dimensions. Thus the coupling constant of the gravitational interaction is no longer the Planck mass $M_P$, because gravity is spread into an additional dimension. This new energy scale $M_{4+1}$ sets a cross-over length $\rc=\frac{M_P^2}{2{M_{4+1}}^3}$ that represents the scale at which the true, five dimensional gravitational effects become dominant with respect to the Einsteinian four dimensional gravity.

The model is constrained by putting all the matter and energy species on the brane, and by letting the fifth dimension be filled only by gravity; the Einstein equations thus give a different relation for the evolution of the scale factor $a(t)=A(t,y=0)$ on the brane. The standard Hubble parameter is $H=\dot a/a$, where a dot denotes a derivative with respect to the proper time, and the expansion history as a function of the redshift $z$ reads
\begin{equation}
\left(\frac{H(z)}{\ho}\right)^2=\left(\frac{1}{2\rc\ho}+\sqrt{\oi\left(1+z\right)^{n_i}+\os+\frac{1}{4{\rc}^2{\ho}^2}}\right)^2+\Omega_K\left(1+z\right)^2,
\end{equation}
where $i=K,m,r$ stands for curvature, matter or radiation, with $n_i=2,3,4$, respectively, and $\os=8\pi G\sigma/\left(3\ho^2\right)$ is a cosmological-constant term which can be interpreted as a non-vanishing tension $\sigma$ of the brane. We expect that, in the higher-dimensional framework, $\os$ could find a natural origin.

Observations indicate that the Universe is spatially flat \cite{Spergel:2006hy}; therefore, the DGP model has been deeply investigated in the case of $\ok\equiv1-\om-\orad-\os-1/(\rc\ho)=0$. In this case, the modified Friedmann equation is
\begin{equation}
H^2\pm\frac{H}{\rc}=\ho^2\left(\oi a^{-n_i}+\os\right).\label{dgp-friedmann}
\end{equation}
Hereafter, we will refer to the upper sign in the modified Friedmann equation as the ``normal'' branch, and to the lower sign as the ``self-accelerating'' branch. Indeed, with the choice of the minus sign, the DGP model can achieve an accelerated expansion without a cosmological constant, unlike the case of the normal branch, where a non-null $\os$ term is needed.

In this paper we focus on a phenomenological extension of DGP model \cite{Dvali:2003rk,Dvali:2006su,Schaefer:2007nf,Afshordi:2008rd,Kobayashi:2009da}. Here, the modified Friedmann equation (\ref{dgp-friedmann}) is generalised as
\begin{equation}
H^2\pm\frac{H^{2\alpha}}{\rc^{2\left(1-\alpha\right)}}=\ho^2\left(\oi a^{-n_i}+\os\right),\label{edgp-friedmann}
\end{equation}
where $\alpha$ is a free parameter which is strictly related to the graviton propagator \cite{Dvali:2003rk,Afshordi:2008rd}. The condition $\ok=0$ clearly implies
\begin{equation}
\os=1-\om-\orad\pm\left(\rc\ho\right)^{2\left(\alpha-1\right)}.
\end{equation}

Such a model presents a graviton propagator which is proportional to
\begin{equation}
\frac{1}{k^2+k^{2\alpha}\rc^{2\left(\alpha-1\right)}}.
\end{equation}
This would follow from a phenomenological model of modified gravity proposed in \cite{Dvali:2006su} and it is a power-law generalisation of the graviton propagator in the DGP brane-world. Since we are interested in a long distance modification of gravity, we assume that $\alpha<1$. The unitarity constraint requires $\alpha\geq0$ \cite{Dvali:2006su}. It can be seen that $\alpha=1$ can be absorbed into a redefinition of the Newtonian gravitational constant $G$. The propagator with $\alpha\ll1$ has some connection to the so-called cascading DGP brane-world \cite{deRham:2007xp} and thus the Friedmann equation with $\alpha\ll1$ might be obtained in such higher codimension models. However, we would like to stress that a detailed analysis of higher codimension DGP models has yet to be undertaken and no brane-world model has been known so far that leads to Eq.~(\ref{edgp-friedmann}). Therefore, we shall view Eq.~(\ref{edgp-friedmann}) as a phenomenological starting point for our modified gravity. It is easy to see that the standard DGP model is recovered when $\alpha=1/2$, while the choice $\alpha=0$, often called ``degravitation,'' leads to an expansion history identical to that of the $\Lambda$CDM model.

\section{Parameter estimation}\label{estimation}
We now estimate the model parameters required to reproduce the observed expansion history of the Universe. 
In the Bayesian approach to model testing, we explore the parameter space through the posterior probability density function
\begin{equation}
\pi(h)L_\mathrm{SNIa}(\mathbf p)L_\mathrm{GRB}(\mathbf p)L_\mathrm{BAO}(\mathbf p)L_\mathrm{WMAP5}(\mathbf p),\label{likelihood}
\end{equation}
where $\mathbf p$ denotes the set of model parameters, $\pi(h)$ is the prior on the Hubble constant $h$ and $L_D$ is the likelihood function related to the dataset $D$.

The Gaussian prior
\begin{equation}
\pi(h)\propto e^{-\frac{1}{2}\left(\frac{h_\mathrm{HST}-h}{\sigma_\mathrm{HST}}\right)^2}
\end{equation}
stems from the results of the HST Project \cite{Freedman:2000cf} which estimated the Hubble constant $\ho$ using a well calibrated set of local distances. Averaging over the different methods, the survey gives:
\begin{equation}
h_\mathrm{HST}\pm\sigma_\mathrm{HST}=0.72\pm0.08
\end{equation}
which is independent of the cosmological model.

The second and the third terms in the likelihood (\ref{likelihood}) are both related to the Hubble diagram, 
and involve SNeIa and GRBs. The SNeIa provided the first piece of evidence of the cosmic acceleration based on the predicted distance modulus
\begin{equation}
\mu_\mathrm{th}(z|\mathbf p)=25+5\log\left[(1+z)\chi(z|\mathbf p)\right],
\end{equation}
where $\de\chi=\de z/H(z)$ is the differential radial comoving distance. The likelihood function is then defined as
\begin{equation}
L_\mathrm{SNIa}(\mathbf p)=\frac{e^{-\frac{\Delta\mu C_\mathrm{SNIa} ^{-1}\Delta\mu^T}{2}}}{\sqrt{\det\left(2\pi C_\mathrm{SNIa}\right)}},
\end{equation}
where $\Delta\mu$ is a $\mathcal N_\mathrm{SNIa}$-vector containing the values $\mu_\mathrm{obs}(z_i)-\mu_\mathrm{th}(z_i)$ and $C_\mathrm{SNIa}$ is the $\mathcal N_\mathrm{SNIa}\times\mathcal N_\mathrm{SNIa}$ covariance matrix of the SNeIa data. Here, $\mathcal N_\mathrm{SNIa}$ is the total number of SNeIa used.

SNeIa are limited to $z\sim1.5$. Therefore we have to resort to a different distance indicator to probe the Hubble diagram to higher redshift. Thanks to the enormous energy release that makes them visibile up to $z\sim6.6$, GRBs stand out as possible candidates for this purpose. As a result, Sch\"afer \cite{Schaefer:2006pa} has provided the first GRB Hubble diagram containing 69 objects with $\mu_\mathrm{obs}(z)$ estimated by averaging over 5 different 2D correlations. Here we use the updated GRB Hubble diagram recently presented in \cite{Cardone:2009mr} based on a model-independent recalibration of the same 2D correlations used by Sch\"afer. Since there is no correlation between the errors of different GRBs, the likelihood function simply reads
\begin{equation}
L_\mathrm{GRB}(\mathbf p)\propto e^{-\frac{1}{2}\sum_{i=1}^{\mathcal N_\mathrm{GRB}}\left[\frac{\mu_\mathrm{obs}(z_i)-\mu_\mathrm{th}(z_i|\mathbf p)}{\sqrt{\sigma_i^2+\sigma_\mathrm{GRB}^2}}\right]^2}
\end{equation}
where $\mathcal N_\mathrm{GRB}$ is the total number of GRBs and $\sigma_\mathrm{GRB}$ quantifies the intrinsic scatter inherited from the scatter of GRBs around the 2D correlations used to derive the individual distance moduli.

$L_\mathrm{BAO}(\mathbf p)$ is the likelihood function related to the BAO acoustic scale, which is effectively constrained by the detected peak in the correlation function of the luminous red galaxies. If we define the effective distance \cite{Eisenstein:2005su,Tegmark:2006az}
\begin{equation}
d_V(z)=\left[\chi^2(z)\frac{z}{H(z)}\right]^\frac{1}{3} \; ,
\end{equation}
the ratio $d_z=r_s(z_d)/d_V(z)$, where $r_s(z)$ is the sound horizon at the drag epoch $z_d$, is a well constrained quantity for $z=0.2$ and $z=0.35$ and its values are $d_{0.2}=0.1905$ and $d_{0.35}=0.1097$ \cite{Eisenstein:2005su,Tegmark:2006az,Komatsu:2008hk}.

Finally, the last term in the likelihood (\ref{likelihood}) is related to the properties of the CMB. Following \cite{Komatsu:2008hk}, we define three fitting parameters for comparison with the WMAP5 data, i.e. the redshift of the surface of last scattering $z_\mathrm{rec}$, the shift parameter at that epoch
\begin{equation}
R(z_\mathrm{rec})=\sqrt{\om}\ho(1+z_\mathrm{rec})d_A(z_\mathrm{rec}),
\end{equation}
and the acoustic scale
\begin{equation}
l_a=(1+z_\mathrm{rec})\frac{\pi d_A(z_\mathrm{rec})}{r_s(z_\mathrm{rec})}.
\end{equation}
Here, $d_A(z)=\chi(z)/(1+z)$ is the proper angular diameter distance. We compare our results with the values $z_\mathrm{rec}=1090.04$, $R(z_\mathrm{rec})=1.71$ and $l_a=302.1$.

The constraints on the individual parameters of the $\Lambda$CDM, eDGP normal and self-accelerating branches (eDGPn and eDGPs, respectively) are summarized in Tables~1-3, where we give the best-fit, mean and median values and the $68\%$ and $95\%$ confidence limits. These constraints are obtained for the entire set of cosmological observational data we choose for this work, i.e. SNeIa, $H(z)$ data, BAOs and GRBs.
\TABULAR[ht!]{cccccc}{\multicolumn{6}{c}{$\Lambda$CDM}\\$x$ & $x_\mathrm{best\,fit}$ & $\langle x\rangle$ & $x_\mathrm{med.}$ & $68\%$ CL & $95\%$ CL\\
\hline
$\om$ & $0.28$ & $0.28$ & $0.28$ & $(0.27,\,0.29)$ & $(0.255,\,0.31)$\\
$10^4\orad$ & $1.1$ & $1.1$ & $1.1$ & $(1.0,\,1.1)$ & $(9.9,\,1.1)$\\
$h$ & $0.71$ & $0.71$ & $0.71$ & $(0.69,\,0.72)$ & $(0.67,\,0.74)$\\
$\sigma_\mathrm{GRB}$ & $0.40$ & $0.37$ & $0.38$ & $(0.27,\,0.47)$ & $(0.10,\,0.57)$}{Constraints from SNeIa, $H(z)$, GRBs and BAOs on the $\Lambda$CDM model parameters.}
\TABULAR[ht!]{cccccc}{\multicolumn{6}{c}{eDGPn}\\
$x$ & $x_\textrm{best-fit}$ & $\langle x\rangle$ & $x_\mathrm{med.}$ & $68\%$ CL & $95\%$ CL\\
\hline
$\om$ & $0.28$ & $0.28$ & $0.28$ & $(0.27,\,0.29)$ & $(0.25,\,0.31)$\\
$10^4\orad$ & $1.1$ & $1.1$ & $1.1$ & $(1.0,\,1.1)$ & $(9.9,\,1.1)$\\
$\rc\ho$ & $52.375$ & $155.04$ & $87.16$ & $(28.20,\,284.94)$ & $(5.72,\,766.268)$\\
$\alpha$ & $0.098$ & $0.116$ & $0.083$ & $(0.033,\,0.187)$ & $(0.013,\,0.481)$\\
$h$ & $0.71$ & $0.71$ & $0.71$ & $(0.69,\,0.72)$ & $(0.67,\,0.74)$\\
$\sigma_\mathrm{GRB}$ & $0.42$ & $0.375$ & $0.38$ & $(0.28,\,0.48)$ & $(0.18,\,0.56)$}{Constraints from SNeIa, $H(z)$, GRBs and BAOs on the eDGPn model parameters.}
\TABULAR[ht!]{cccccc}{\multicolumn{6}{c}{eDGPs}\\
$x$ & $x_\textrm{best-fit}$ & $\langle x\rangle$ & $x_\mathrm{med.}$ & $68\%$ CL & $95\%$ CL\\
\hline
$\om$ & $0.28$ & $0.28$ & $0.28$ & $(0.27,\,0.295)$ & $(0.25,\,0.31)$\\
$10^4\orad$ & $1.1$ & $1.1$ & $1.1$ & $(1.0,\,1.1)$ & $(9.9,\,1.15)$\\
$\rc\ho$ & $19.43$ & $88.245$ & $73.33$ & $(24.20,\,146.10)$ & $(3.225,\,287.58)$\\
$\alpha$ & $0.517$ & $0.103$ & $0.077$ & $(0.029,\,0.166)$ & $(0.013,\,0.400)$\\
$h$ & $0.71$ & $0.71$ & $0.71$ & $(0.69,\,0.72)$ & $(0.67,\,0.74)$\\
$\sigma_\mathrm{GRB}$ & $0.415$ & $0.36$ & $0.37$ & $(0.25,\,0.48)$ & $(0.90,\,0.56)$}{Constraints from SNeIa, $H(z)$, GRBs and BAOs on the eDGPs model parameters.}

\begin{figure}[ht!]
\centering
\includegraphics[width=0.9\textwidth]{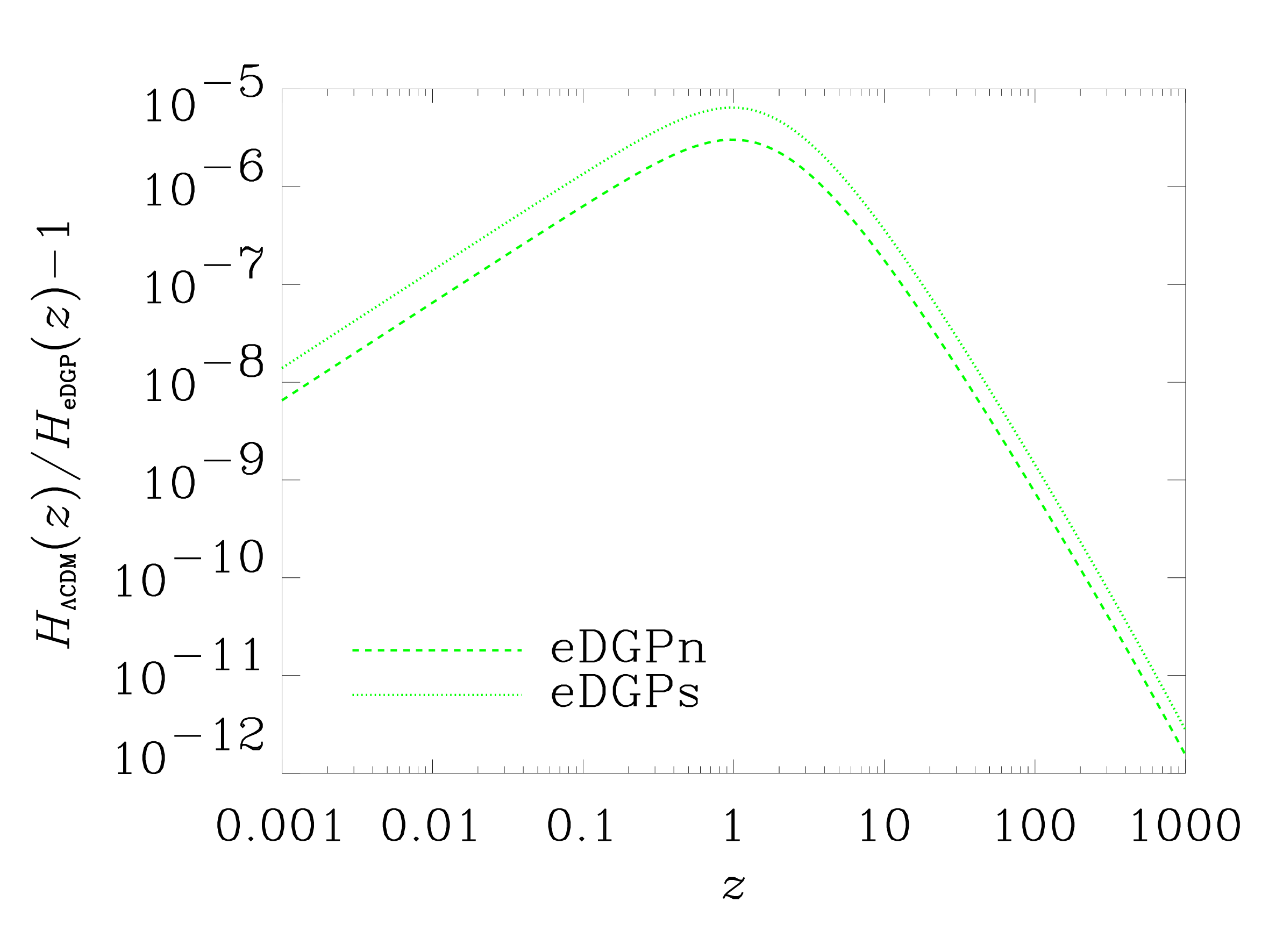}
\caption{Relative error between the $\Lambda$CDM and the eDGP expansion histories.}\label{w_DGP}
\end{figure}

We find that the eDGP model is able to fit these datasets. Moreover, in Fig.~\ref{w_DGP} we show how the expansion rates in $\Lambda$CDM and in the eDGP model differ by less than $0.001\%$ over a wide range of redshift.

\subsection{CMB anisotropy power spectrum}\label{cmb}
The CMB probes the geometry of the background expansion as well as the formation of LSS. To predict these effects we use the parameterised-post Friedmannian \cite{Hu:2007pj} modifications to the \textsc{camb} (Code for Anisotropies in the Microwave Background) routines\footnote{http://camb.info/} \cite{Lewis:1999bs,Fang:2008sn,Fang:2008kc}.

We perform a chi-squared analysis with the most recent data coming from WMAP7 \cite{Larson:2010gs} to get the best-fit value for the temperature anisotropy power spectrum normalisation, for the $\Lambda$CDM model and both branches of the eDGP model. We thus obtain the normalisation of the matter power spectrum that we use in the next section and in further calculations. The three fits of our models to the CMB data are indistinguishable: with the WMAP7 best-fit value $n_s=0.963$ for the tilt of the primordial power spectrum, the three reduced chi-squared are $\bar\chi^2_\textrm{eDGPn}=\bar\chi^2_\textrm{eDGPs}=\bar\chi^2_\textrm{$\Lambda$CDM}=1.157$. We derive  $\sigma_8=0.858$, the rms mass fluctuations on the scale $R=8\,h^{-1}\mathrm{Mpc}$.

\section{Power spectra}\label{observables}

\subsection{Matter power spectrum}\label{matterpower}
A faster growth of structure at early times, followed by a slower structure growth from the point on where the modification of gravity becomes important, is a feature common to eDGP models and several modified gravity theories. As usual, we define the growth factor $D_+(a)=\delta(\mathbf x,a)/\delta(\mathbf x,a=1)$, where $\delta=\delta\rho/\rho$ is the density contrast. The evolution equation of $D_+$ in the matter-dominated epoch for DGP brane-worlds becomes \cite{Lue:2002sw,Lue:2003ky,Lue:2004rj,Koyama:2005kd,Koyama:2006ef,Schaefer:2007nf}
\begin{equation}
\ddot\delta+2H\dot\delta=\frac{3}{2}\om\ho^2\left(1+\frac{1}{3\beta}\right)\frac{\delta}{a^3},\label{growtheq-dgp}
\end{equation}
where $\delta=\delta\rho/\rho$ is the density contrast and
\begin{equation}
\beta=1\pm2\left(\rc H\right)^{2\left(1-\alpha\right)}\left[1+\frac{2\left(1-\alpha\right)\dot H}{3H^2}\right].
\end{equation}

We have shown that the Hubble parameter in both branches of the eDGP model differs by less than $0.001\%$ from that of the $\Lambda$CDM model (Fig.~\ref{w_DGP}), and the normalisation of the primordial power spectrum obtained from the CMB temperature anisotropy analysis is again the same in both eDGPs and eDGPn (\S~\ref{cmb}). Therefore, hereafter our results will refer only to the eDGP model, because the curves in the two branches are almost indistinguishable.

\subsubsection{Linear r\'egime}
By performing the Fourier transform $\delta_k(a)$ of the density contrast, we can construct the present-day matter power spectrum $P(k)\equiv P^\delta\left(k,a=1\right)$, where, for a generic field $f_k(a)$
\begin{equation}
\langle f_k(a){f_{k'}}^\ast(a)\rangle=\left(2\pi\right)^3\delta_D\left(k-k'\right)P^f(k,a)\label{P(k)},
\end{equation}
with $\delta_D$ the Dirac $\delta$ function. In the DGP model, the Newtonian potential $\Phi_k(a)$ and the metric potential $\Psi_k(a)$ depend on the scale $k$ and thus their evolution deeply differs from the corresponding quantities in the $\Lambda$CDM model; however, the growth factor does evolve similarly to $\Lambda$CDM for a wide range of scales \cite{Cardoso:2007xc}. Thanks to this peculiarity, we can use the growth factor $D_+(a)$, which is solution of Eq.~(\ref{growtheq-dgp}), to construct the matter power spectrum as
\begin{equation}
P^\delta(k,z)=2\pi^2{\delta_H}^2\left(\frac{k}{\ho^3}\right)^{n_s}T^2(k)\left[\frac{D_+(z)}{D_+(z=0)}\right]^2.
\end{equation}
Here, ${n_s}$ is the tilt of the primordial power spectrum, $\delta_H$ is its normalisation and $T(k)$ is the matter transfer function which describes the evolution of perturbations through the epochs of horizon crossing and radiation-matter transition. As mentioned in \S~\ref{cmb}, we use $n_s=0.963$ and $\delta_H$ such that $\sigma_8=0.858$.
The transfer function is in Ref.~\cite{Eisenstein:1997jh}.

\subsubsection{Non-linear r\'egime}
To obtain the non-linear matter power spectrum, whose contribution is particularly relevant for computing the weak lensing cosmic shear signal, we use two different prescriptions. On one hand, $N$-body simulations in eDGP models and degravitation theories have recently been performed \cite{Khoury:2009tk} and they show that, in order to correctly reproduce the matter power spectrum, the \textsc{halofit} fitting formul\ae\ \cite{Smith:2002dz} have to be slightly modified: the new set of parameters works well for the whole range of $\alpha$, although additional $N$-body simulations covering a wider volume in the parameter space still lack. The new parameters of the fitting formul\ae\ are listed in Table~\ref{KW-params}. The parameter $\eth$ is related to a redefinition of the $y$ parameter of \textsc{halofit}, i.e. $y\rightarrow\eth y$. We will denote this non-linear perscription as ``KW'' from Khoury and Wyman (2009), Ref.~\cite{Khoury:2009tk}. 
\begin{table}[ht!]
\centering
\begin{tabular}{cc}
$\log_{10}a_n$&$0.84\log_{10}a^\mathrm{std}_n$\\
$\log_{10}b_n$&$\log_{10}b^\mathrm{std}_n+\log_{10}1.1$\\
$\log_{10}c_n$&$\log_{10}c^\mathrm{std}_n+\log_{10}1.05$\\
$\log_{10}\mu_n$&$\log_{10}\mu^\mathrm{std}_n+\log_{10}0.875$\\
$\log_{10}\nu_n$&$\log_{10}\nu^\mathrm{std}_n+\log_{10}0.875$\\
$\alpha_n$&$0.8\alpha^\mathrm{std}_n$\\
$\beta_n$&$1.95\beta^\mathrm{std}_n$\\
\hline
$\eth$&$1.035$
\end{tabular}
\caption{Parameters for the modified non-linear fitting algorithm from \protect\cite{Khoury:2009tk} (left column) as a function of the parameters in the standard \textsc{halofit} formula \protect\cite{Smith:2002dz} (right column).}\label{KW-params}
\end{table}

For modified gravity to agree with Solar system observations, the non-linear matter power spectrum has to approach the standard $\Lambda$CDM solution on small scales. This means that the non-linear power spectrum has to be an interpolation of two power spectra. The first is the modified gravity non-linear power spectrum $P^\delta_\mathrm{MG}(k,z)$, which is obtained without the non-linear interactions that are responsible for the recovery of GR. This is equivalent to assume that gravity is modified down to small scales in the same way as in the linear r\'egime. The second term, $P^\delta_\mathrm{GR}(k,z)$, is the non-linear power spectrum obtained in the DE model that follows the same expansion history of the Universe as the modified gravity model, yet obeying to GR. In other words, this is the non-linear $P^\delta(k,z)$ which will have a $\Lambda$CDM model with an expansions history $H(z)$ equivalent to that of the modified gravity theory. Since the Hubble parameter in the eDGP model with the parameter values obtained with our Markov Chains is the same as that in the $\Lambda$CDM model, $P^\delta_\mathrm{GR}(k,z)$ is simply the $\Lambda$CDM matter power spectrum. We use the fitting formula proposed by Hu and Sawicki (2007), Ref.~\cite{Hu:2007pj}
\begin{equation}
P^\delta_\mathrm{nl}(k,z)=\frac{P^\delta_\mathrm{MG}(k,z)+c_\mathrm{nl}(z)\Sigma^2(k,z)P^\delta_\mathrm{GR}(k,z)}{1+c_\mathrm{nl}(z)\Sigma^2(k,z)},
\end{equation}
where $\Sigma^2(k,z)=\left[k^2P^\delta(k,z)/2\pi^2\right]^{a_1}$ picks out non-linear scales, since $P^\delta(k,z)$ is the linear power spectrum of Eq.~(\ref{P(k)}); $c_\mathrm{nl}(z)=A\left(1+z\right)^{a_2}$ determines the scale at which the power spectrum approaches the GR result as a function of redshift. Their functional forms have been obtained by perturbation theory \cite{Koyama:2009me} and confirmed by $N$-body simulations \cite{Oyaizu:2008tb,Schmidt:2009sg}. Here, we use $a_1=1$, $a_2=0.16$ and $A=0.3$. We will denote this non-linear perscription as ``HS.''

In addition to the two non-linear prescriptions mentioned above, Schmidt et al. \cite{Schmidt:2009yj} have recently studied the spherical collapse model in the DGP brane-world gravity. They show how results on the mass function, halo bias and power spectrum obtained with $N$-body simulations \cite{Schmidt:2009sg,Schmidt:2009sv} can be described semi-analytically. However, they find a tension between the simulations and their reconstructed power spectrum. Specifically, the $P_\mathrm{nl}(k)$ in the self-accelerating branch matches the simulated data very well, whereas there is a $10\div30\%$ discrepancy in the normal branch. Nonetheless, their spherical collapse predictions appear to be able to provide an effective way of including non-linear effects in results obtained from DGP simulations in the linear r\'egime. Further investigations are worth performing, in particular in the perspective of generalising such an approach to the eDGP model.

\begin{figure}[ht!]
\centering
\includegraphics[width=0.9\textwidth]{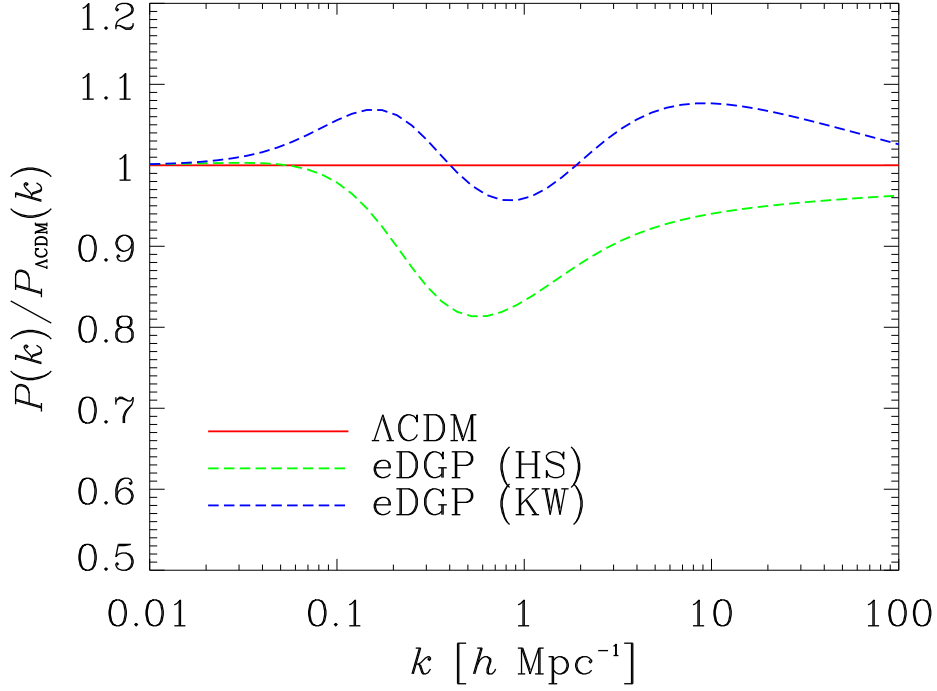}
\caption{Ratio of the non-linear matter power spectra of the $\Lambda$CDM and the eDGP models. For the eDGP model, the non-linear power spectra obtained with \cite{Hu:2007pj,Khoury:2009tk} are shown. }\label{matter_power_spetrum-nl}
\end{figure}

In Fig.~\ref{matter_power_spetrum-nl} we show the non-linear matter power spectra for the $\Lambda$CDM and the eDGP models. For the latter, we show the power spectra obtained with both the KW and the HS methods described above. As expected, the eDGP-HS non-linear $P(k)$ approaches the $\Lambda$CDM signal at large $k$'s, because the prescription was explicitly proposed to track the non-linear behaviour of GR at small scales. Unexpectedly, we find that the KW prescription also is able to reproduce the non-linear $\Lambda$CDM power spectrum. This result was not found in \cite{Khoury:2009tk}, because our $\alpha$ and $\rc$, unlike theirs, derive from a Markov Chain Monte Carlo analysis which is based on the data describing the expansion history of the real Universe.

\subsection{Weak lensing signal}\label{weaklensing}
It is known from GR that light beam paths are curved by the presence of matter. In the weak lensing framework the deflection of light is small and, consequently, we can use Born's approximation, where lensing effects are evaluated on the null-geodesic of the unperturbed (unlensed) photon \cite{Hu:2000ee,Bartelmann:1999yn}.

In the DGP model, and similarly in its phenomenological extensions, for the two potentials $|\Phi|\neq|\Psi|$ holds. Thus, the relation between the distribution of matter overdensities in the Universe and the two metric perturbations, the potentials $\Phi$ and $\Psi$, is not trivial. In GR, in the matter-dominated era, when there is no anisotropic stress, $\Phi=-\Psi$ and therefore we can simply use the Newtonian potential $\Phi$, thanks to the canonical Poisson equation, to compute cosmic convergence and shear. However, in general, the weak lensing effect is due to the combination of both the Newtonian and the metric potential. We will refer to this combination as the ``deflecting potential,'' and we will denote it with\footnote{In the literature, the combinations of the two potentials are often indicated with $\Phi_\pm=-\left(\Phi\pm\Psi\right)/2$. However, different authors use to refer to the metric potentials in the length element differently. For the sake of simplicity, we have chosen this notation, because we are only interested in the combined effect responsible for the weak lensing signal.}
\begin{equation}
\Upsilon\equiv-\frac{\Phi-\Psi}{2}.
\end{equation}

A technique for obtaining the Fourier modes of the deflecting potential starting from the growth factor in the extended DGP model has recently been proposed \cite{Khoury:2009tk}. This modified Poisson equation reads
\begin{equation}
\Upsilon_k(a)=\frac{3}{2}\om\ho^2\frac{\mathscr G(k,a)}{G}\frac{\sqrt{P^\delta(k,a)}}{k^2a},
\end{equation}
where
\begin{equation}
\mathscr G(k,a)=\frac{G}{1+\left(\frac{k\rc}{a}\right)^{2\left(\alpha-1\right)}}\frac{1-\frac{2g}{\epsilon}\left(\sqrt{1+\epsilon}-1\right)}{1-g}
\end{equation}
and
\begin{align}
\epsilon(k,a)&=8g^2H^2\rc^2\om\delta_k,\\
g&=-\frac{1}{3\beta}.
\end{align}
$P^\delta(k,a)$ is the matter power spectrum as a function of the comoving wavenumber $k$ and the scale factor $a$, as computed in \S~\ref{matterpower}.

All weak lensing observables may be expressed in terms of the projected potential
\begin{equation}
\phi(\hat{\mathbf n})=\int\!\!\de\chi\,\frac{W(\chi)}{\chi^2}\Upsilon(\hat{\mathbf n},\chi)\label{phi}
\end{equation}
where
\begin{equation}
W(\chi)=-\chi\int_\chi^\infty\de\chi'\,\frac{\chi'-\chi}{\chi'}n(\chi')\label{W(z)}
\end{equation}
is the weight function of weak lensing, with $n\left[\chi(z)\right]$ representing the redshift distribution of the sources, such that $\int\!\!\de\chi\,n(\chi)=1$. We now introduce a distortion tensor \cite{Kaiser:1996tp,Bartelmann:1999yn}
\begin{equation}
\phi_{,ij}(\mathbf x)=\int_0^\chi\!\!\de\chi'\,\chi'W(\chi')\Upsilon_{,ij}(\hat{\mathbf n},\chi')\label{phi,ij},
\end{equation}
where commas denote derivatives with respect to directions perpendicular to the line of sight. The trace of the distortion tensor represents the cosmic convergence
\begin{equation}
\kappa(\mathbf x)=\phi_{,11}(\mathbf x)+\phi_{,22}(\mathbf x)
\end{equation}
and the (complex) shear is the linear combination
\begin{equation}
\gamma(\mathbf x)=\phi_{,11}(\mathbf x)-\phi_{,22}(\mathbf x)+2i\phi_{,12}(\mathbf x).
\end{equation}

In the flat-sky approach we expand the projected potential $\phi(\hat{\mathbf n})$ in its Fourier modes
\begin{equation}
\phi(\hat{\mathbf n})=\int\!\!\frac{\de^2\ell}{{(2\pi)}^2}\,\phi(\boldsymbol{\ell})e^{i\boldsymbol{\ell}\cdot\hat{\mathbf n}}.
\end{equation}
The power spectrum is defined as the Fourier transform of the 2D correlation function
\begin{equation}
\langle\phi^\ast(\boldsymbol{\ell})\phi(\boldsymbol{\ell}')\rangle={(2\pi)}^2\delta_D(\boldsymbol{\ell}-\boldsymbol{\ell}')C^{\phi\phi}(\ell).
\end{equation}
Thus, we have \cite{Kaiser:1991qi}
\begin{equation}
C^{\phi\phi}(\ell)=\int_0^\infty\!\!\de\chi\,\frac{W^2(\chi)}{\chi^6}P^\Upsilon\left(\frac{\ell}{\chi},\chi\right)\label{C(l)}
\end{equation}
where $P^\Upsilon(k,z)$ is the power spectrum of the deflecting potential constructed following Eq.~(\ref{P(k)}), and we have introduced Limber's approximation, where the only Fourier modes that contribute to the integral are those with $\ell\gg k\chi$. The cosmic convergence and shear can be obtained with
\begin{equation}
C^{\kappa\kappa}(\ell)=C^{\gamma\gamma}(\ell)\equiv\ell^4C^{\phi\phi}(\ell).
\end{equation}

We compute the shear power spectrum of light emitted by background galaxies, because this effect is relevant for the study of alternative/modified gravity theories. It has been shown that, although the CMB convergence can give a signal similar to that of the $\Lambda$CDM model \cite{Camera:2009uz}, the background galaxy power spectra can be different. This is due to the simple fact that models alternative to the standard $\Lambda$CDM model have to reproduce the same behaviour in the early Universe, where CMB constraints are stricter, but can differ at late times. 

\begin{figure}[ht!]
\centering
\includegraphics[width=0.9\textwidth]{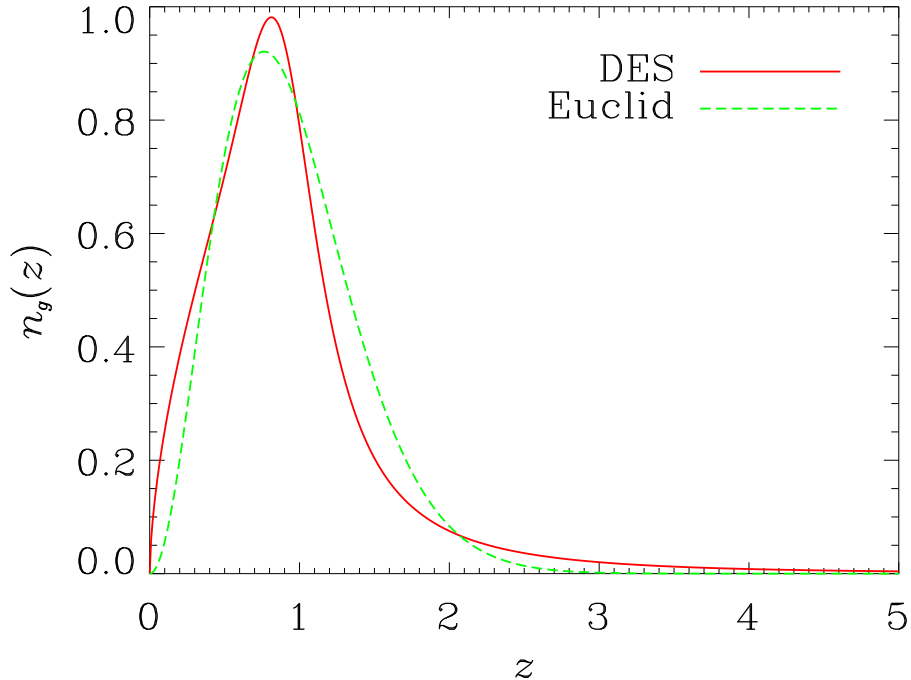}
\caption{The source redshift distributions used: for ground-based surveys (DES) with median redshift $z_m=0.826$ and for a space-survey (Euclid) with median redshift $z_m=0.91$.}\label{des}
\end{figure}

We calculate results for realistic oncoming surveys: a ground-based survey similar to DES, and a space-based survey such as Euclid, using the redshift distributions shown in Fig.~\ref{des}.
The redshift distribution for our ground-based survey is chosen to be the same as for CFHTLS \cite{Fu:2007qq} giving a median redshift $z_m=0.826$, a sky coverage $\Theta_\mathrm{deg}^2=5,000\,\mathrm{deg}^2$ and a mean galaxy number density $\bar n=13.3\,\mathrm{arcmin}^{-2}$; for Euclid we use the distribution given by \cite{Blake:2004tr,Hawken:2009fr} with $z_m=0.91$, $\Theta_\mathrm{deg}^2=20,000\,\mathrm{deg}^2$ and $\bar n=35\,\mathrm{arcmin}^{-2}$.

\begin{figure}[ht!]
\centering
\includegraphics[width=0.9\textwidth]{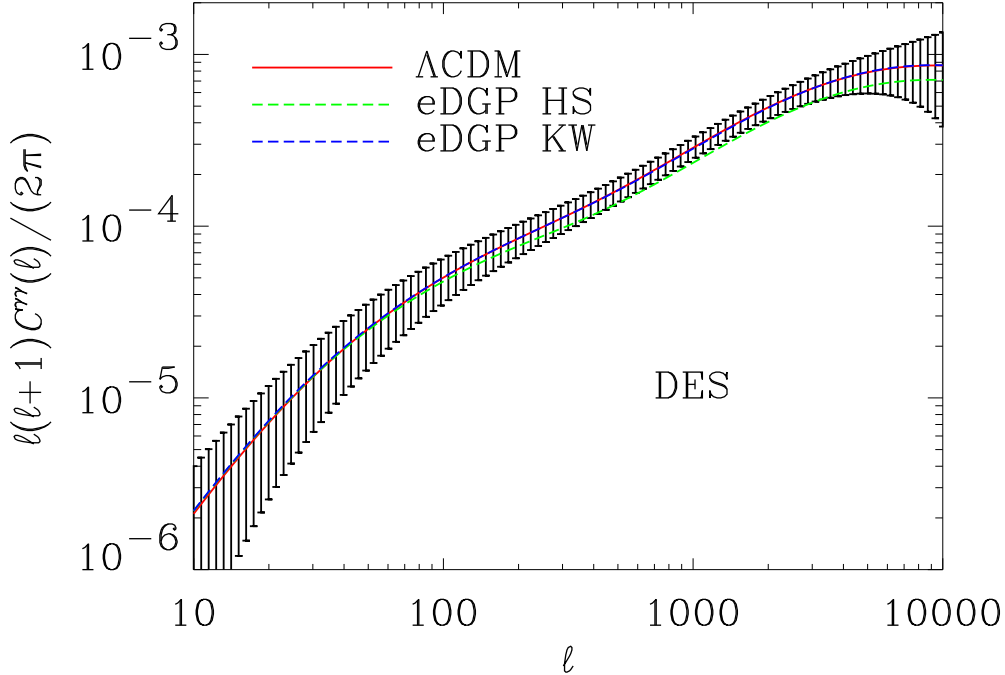}
\caption{Shear power spectra $\ell(\ell+1)C^{\gamma\gamma}(\ell)/(2\pi)$ in the eDGP model and in the standard $\Lambda$CDM model of photons coming from background galaxies expected from a ground-based DES-like survey distribution of sources with median redshift $z_m=0.826$: blue/dark with the KW non-linear prescription \protect\cite{Khoury:2009tk}, green/light with HS \protect\cite{Hu:2007pj}. Error bars are $1\sigma$.}\label{convergence_power_spetrum-bg-DES}
\end{figure}

\begin{figure}[ht!]
\centering
\includegraphics[width=0.9\textwidth]{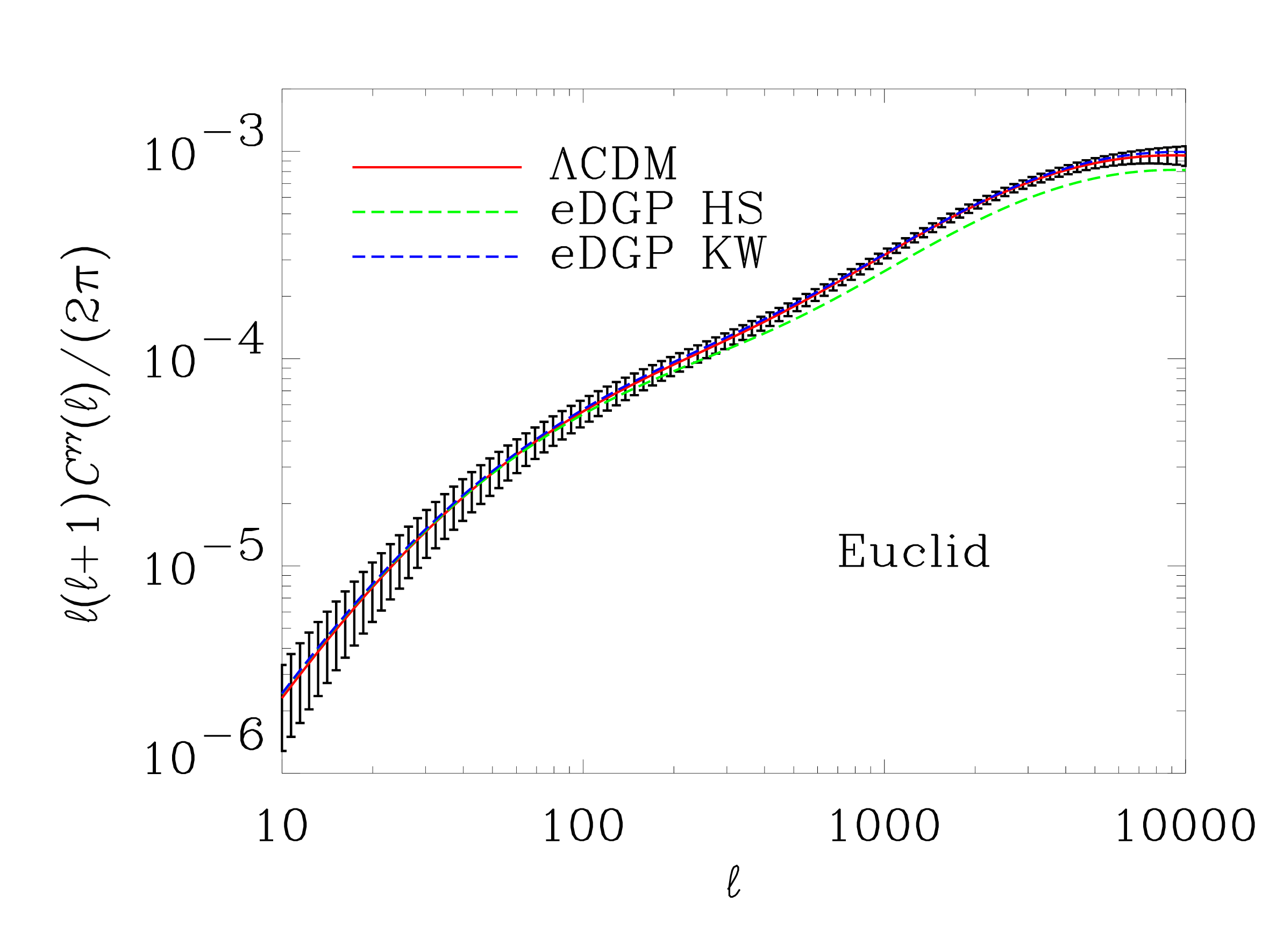}
\caption{Shear power spectra $\ell(\ell+1)C^{\gamma\gamma}(\ell)/(2\pi)$ in the eDGP model and in the standard $\Lambda$CDM model of photons coming from background galaxies expected from a space-based Euclid-like survey distribution of sources with median redshift $z_m=0.91$: blue/dark with the KW non-linear prescription \protect\cite{Khoury:2009tk}, green/light with HS \protect\cite{Hu:2007pj}. Error bars are $1\sigma$.}\label{convergence_power_spetrum-bg-Euclid}
\end{figure}

Figs.~\ref{convergence_power_spetrum-bg-DES}-\ref{convergence_power_spetrum-bg-Euclid} show the shear power spectra $\ell(\ell+1)C^{\gamma\gamma}(\ell)/(2\pi)$ of light from background galaxies in the eDGP model and in the standard $\Lambda$CDM model. The shear signal shows the same behaviour of the non-linear matter power spectrum shown in Fig.~\ref{matter_power_spetrum-nl}: the weak lensing signal in eDGP model appears to be suppressed with respect to $\Lambda$CDM, when using the HS procedure for computing the non-linear $P^\delta(k,z)$, whereas the eDGP-KW shear signal is virtually indistinguishable from $\Lambda$CDM.

We also show the expected errors as computed with \cite{Kaiser:1991qi,Kaiser:1996tp}
\begin{equation}
\Delta C^{\gamma\gamma}(\ell)=\sqrt{\frac{2}{2(\ell+1)f_\mathrm{sky}}}\left(C^{\gamma\gamma}(\ell)+\frac{\langle{\gamma_\mathrm{int}}^2\rangle}{\bar n}\right),
\end{equation}
where $f_\mathrm{sky}=\Theta_\mathrm{deg}^2\pi/129,600$ is the fraction of the sky covered by a survey of area $\Theta_\mathrm{deg}^2$ and $\langle{\gamma_\mathrm{int}}^2\rangle^{0.5}\simeq0.4$ is the galaxy-intrinsic rms shear in one component. The errorbars clearly show that the depth and the wide field of a space-based survey is necessary to discriminate between the $\Lambda$CDM and the extension of the DGP model. In fact, DES could do so, in principle, only at $1\sigma$ level in the range of angular scales $500\lesssim\ell\lesssim3000$, whereas, thanks to its wide and deep field, Euclid can achieve $3\sigma$ level in the range $700\lesssim\ell\lesssim3000$.


\section{Conclusions}\label{conclusions}
We investigate a phenomenological extension of the Dvali-Gabadadze-Porrati (DGP) brane-world cosmology, which presents an extra parameter $\alpha$ as an exponent of the non-quadratic term in the modified Friedmann equation. The DGP model admits two Hubble parameter $H(z)$ solutions of the modified Friedmann equations: one of them presents a self-accelerating behaviour at late-times even without any cosmological constant term, hence it is known as the ``self-accelerating'' branch. The other solution is often called the ``normal'' branch. Similarly, the extended DGP model has two expansion histories of the Universe, and we refer to them as eDGPn and eDGPs, for the normal and self-accelerating branch, respectively. In this work, we allow the eDGP model to have an explicit cosmological-constant-like term in the modified Friedmann equation, which can be interpreted as a non-vanishing tension $\sigma$ on the brane.

We compute Monte Carlo Markov Chains to estimate cosmological parameters consistent with current observations. We select a sample of cosmological observables. In particular, we use $H(z)$ data, Type Ia Supernov\ae\ (SNeIa) and Gamma Ray Bursts (GRBs) as standard candles, Baryon Acoustic Oscillations (BAOs) from SDSS and distance priors from WMAP5 data. We run the Markov Chains for both branches of the eDGP model and for the $\Lambda$CDM model, in order to make direct comparisons between the two models. We find that the eDGP models can fit current data at least as well as the standard $\Lambda$CDM model.

By using an extension of the \textsc{camb} code with the parameterised-post Friedmannian framework for modified gravity, we also obtain the CMB temperature anisotropy power spectrum for the eDGP models. With the cosmological parameters obtained with our Markov Chains, we find a perfect agreement between the recent WMAP7 data and the power spectra, for both $\Lambda$CDM and the eDGP models. Specifically, the chi-squared analysis gives $\bar\chi^2_\textrm{eDGPn}=\bar\chi^2_\textrm{eDGPs}=\bar\chi^2_\textrm{$\Lambda$CDM}=1.157$.
The only parameter left free by the Markov Chain Monte Carlo simulations is $\sigma_8$, the rms mass fluctuations in a sphere with radius $8\,h^{-1}\mathrm{Mpc}$. Our chi-squared analysis of the CMB power spectrum 
yields $\sigma_8=0.858$. 

With this normalisation, we compute the matter power spectrum of the density perturbations in the linear r\'egime. When we correct the linear power spectrum by using the recent modification to the \textsc{halofit} fitting formul\ae\ of Ref.~\cite{Khoury:2009tk} (KW) for the eDGP model, we see that, at non-linear scales, unlike the results reported in \cite{Khoury:2009tk}, the matter power spectrum in both branches of the eDGP model approaches the $\Lambda$CDM curve. This result is a consequence of the fact that our $\alpha$ and $\rc$ parameters derive from the Monte Carlo Markov Chains applied to the current data of the Universe expansion history. We also use the non-linear interpolation of Ref.~\cite{Hu:2007pj} (HS), which has been explicitly proposed to track the non-linear behaviour of GR at small scales.

Finally, we compute the weak lensing shear power spectrum as it is expected from a space-based mission such as Euclid and from a ground-based survey like DES, and we show that the eDGP model is difficult to distinguish from $\Lambda$CDM: indeed, at both large and very small angular scales the observational constraints are loose and the differences between the two models are within the uncertainties. Specifically, if the correct non-linear prescription were KW, the $\Lambda$CDM and eDGP shear signal would be degenerate. On the contrary, with the HS non-linear prescription, the $\Lambda$CDM and the eDGP models differ. However, DES is unable to discriminate between the two models because they differ by at most $1\sigma$ in the range of angular scales $500\lesssim\ell\lesssim3000$. On the other hand, the depth and the wide field of a space-based survey such as Euclid can in principle discriminate between the eDGP and the $\Lambda$CDM models, because they differ by $3\sigma$ in the range $700\lesssim\ell\lesssim3000$.

\acknowledgments We thank the referee for a careful reading of our manuscript and very insightful comments. SC and AD gratefully acknowledge partial support from the INFN grant PD51 and the PRIN-MIUR-2008 grant ``Matter-antimatter asymmetry, dark matter and dark energy in the LHC Era.'' SC also wishes to thank Anthony Lewis for his support with the parameterised-post Friedmannian module for \textsc{camb}.

\bibliographystyle{JHEP}
\bibliography{/home/camera/Documenti/LaTeX/Bibliography}

\providecommand{\href}[2]{#2}\begingroup\raggedright\begin{thebibliography}{10}

\bibitem{Zwicky:1933gu}
F.~Zwicky, {\it {Spectral displacement of extra galactic nebulae}},  {\em Helv.
  Phys. Acta} {\bf 6} (1933) 110--127.

\bibitem{Zwicky:1937zza}
F.~Zwicky, {\it {On the Masses of Nebulae and of Clusters of Nebulae}},  {\em
  Astrophys. J.} {\bf 86} (1937) 217--246.

\bibitem{Dodelson:2001ux}
{\bf SDSS} Collaboration, S.~Dodelson {\em et.~al.}, {\it {The
  three-dimensional power spectrum from angular clustering of galaxies in early
  SDSS data}},  {\em Astrophys. J.} {\bf 572} (2001) 140--156,
  [\href{http://xxx.lanl.gov/abs/astro-ph/0107421}{{\tt astro-ph/0107421}}].

\bibitem{Hawkins:2002sg}
E.~Hawkins {\em et.~al.}, {\it {The 2dF Galaxy Redshift Survey: correlation
  functions, peculiar velocities and the matter density of the Universe}},
  {\em Mon. Not. Roy. Astron. Soc.} {\bf 346} (2003) 78,
  [\href{http://xxx.lanl.gov/abs/astro-ph/0212375}{{\tt astro-ph/0212375}}].

\bibitem{Spergel:2006hy}
{\bf WMAP} Collaboration, D.~N. Spergel {\em et.~al.}, {\it {Wilkinson
  Microwave Anisotropy Probe (WMAP) three year results: Implications for
  cosmology}},  {\em Astrophys. J. Suppl.} {\bf 170} (2007) 377,
  [\href{http://xxx.lanl.gov/abs/astro-ph/0603449}{{\tt astro-ph/0603449}}].

\bibitem{Riess:2006fw}
A.~G. Riess {\em et.~al.}, {\it {New Hubble Space Telescope Discoveries of Type
  Ia Supernovae at $z > 1$: Narrowing Constraints on the Early Behavior of Dark
  Energy}},  {\em Astrophys. J.} {\bf 659} (2007) 98--121,
  [\href{http://xxx.lanl.gov/abs/astro-ph/0611572}{{\tt astro-ph/0611572}}].

\bibitem{Hinshaw:2008kr}
{\bf WMAP} Collaboration, G.~Hinshaw {\em et.~al.}, {\it {Five-Year Wilkinson
  Microwave Anisotropy Probe (WMAP) Observations:Data Processing, Sky Maps, \&
  Basic Results}},  \href{http://xxx.lanl.gov/abs/0803.0732}{{\tt
  arXiv:0803.0732}}.

\bibitem{Komatsu:2008hk}
{\bf WMAP} Collaboration, E.~Komatsu {\em et.~al.}, {\it {Five-Year Wilkinson
  Microwave Anisotropy Probe (WMAP) Observations:Cosmological Interpretation}},
   {\em Astrophys. J. Suppl.} {\bf 180} (2009) 330--376,
  [\href{http://xxx.lanl.gov/abs/0803.0547}{{\tt arXiv:0803.0547}}].

\bibitem{Riess:1998cb}
{\bf Supernova Search Team} Collaboration, A.~G. Riess {\em et.~al.}, {\it
  {Observational Evidence from Supernovae for an Accelerating Universe and a
  Cosmological Constant}},  {\em Astron. J.} {\bf 116} (1998) 1009--1038,
  [\href{http://xxx.lanl.gov/abs/astro-ph/9805201}{{\tt astro-ph/9805201}}].

\bibitem{Knop:2003iy}
{\bf Supernova Cosmology Project} Collaboration, R.~A. Knop {\em et.~al.}, {\it
  {New Constraints on $\Omega_M$, $\Omega_\Lambda$, and w from an Independent
  Set of Eleven High-Redshift Supernovae Observed with HST}},  {\em Astrophys.
  J.} {\bf 598} (2003) 102,
  [\href{http://xxx.lanl.gov/abs/astro-ph/0309368}{{\tt astro-ph/0309368}}].

\bibitem{Riess:2004n}
{\bf Supernova Search Team} Collaboration, A.~G. Riess {\em et.~al.}, {\it
  {Type Ia Supernova Discoveries at z>1 From the Hubble Space Telescope:
  Evidence for Past Deceleration and Constraints on Dark Energy Evolution}},
  {\em Astrophys. J.} {\bf 607} (2004) 665--687,
  [\href{http://xxx.lanl.gov/abs/astro-ph/0402512}{{\tt astro-ph/0402512}}].

\bibitem{Larson:2010gs}
D.~Larson {\em et.~al.}, {\it {Seven-Year Wilkinson Microwave Anisotropy Probe
  (WMAP) Observations: Power Spectra and WMAP-Derived Parameters}},  {\em
  arXiv:1001.4635} (2010) [\href{http://xxx.lanl.gov/abs/1001.4635}{{\tt
  arXiv:1001.4635}}].

\bibitem{Bianchi:2010uw}
E.~Bianchi and C.~Rovelli, {\it {Why all these prejudices against a
  constant?}},  \href{http://xxx.lanl.gov/abs/1002.3966}{{\tt
  arXiv:1002.3966}}.

\bibitem{Dvali:2000rv}
G.~R. Dvali, G.~Gabadadze, and M.~Porrati, {\it {Metastable gravitons and
  infinite volume extra dimensions}},  {\em Phys. Lett.} {\bf B484} (2000)
  112--118, [\href{http://xxx.lanl.gov/abs/hep-th/0002190}{{\tt
  hep-th/0002190}}].

\bibitem{Dvali:2000hr}
G.~R. Dvali, G.~Gabadadze, and M.~Porrati, {\it {4D gravity on a brane in 5D
  Minkowski space}},  {\em Phys. Lett.} {\bf B485} (2000) 208--214,
  [\href{http://xxx.lanl.gov/abs/hep-th/0005016}{{\tt hep-th/0005016}}].

\bibitem{Lue:2002sw}
A.~Lue and G.~Starkman, {\it {Gravitational leakage into extra dimensions:
  Probing dark energy using local gravity}},  {\em Phys. Rev.} {\bf D67} (2003)
  064002, [\href{http://xxx.lanl.gov/abs/astro-ph/0212083}{{\tt
  astro-ph/0212083}}].

\bibitem{Tanaka:2003zb}
T.~Tanaka, {\it {Weak gravity in DGP braneworld model}},  {\em Phys. Rev.} {\bf
  D69} (2004) 024001, [\href{http://xxx.lanl.gov/abs/gr-qc/0305031}{{\tt
  gr-qc/0305031}}].

\bibitem{Lue:2004rj}
A.~Lue, R.~Scoccimarro, and G.~D. Starkman, {\it {Probing Newton's constant on
  vast scales: DGP gravity, cosmic acceleration and large scale structure}},
  {\em Phys. Rev.} {\bf D69} (2004) 124015,
  [\href{http://xxx.lanl.gov/abs/astro-ph/0401515}{{\tt astro-ph/0401515}}].

\bibitem{Koyama:2005kd}
K.~Koyama and R.~Maartens, {\it {Structure formation in the DGP cosmological
  model}},  {\em JCAP} {\bf 0601} (2006) 016,
  [\href{http://xxx.lanl.gov/abs/astro-ph/0511634}{{\tt astro-ph/0511634}}].

\bibitem{Cardoso:2007xc}
A.~Cardoso, K.~Koyama, S.~S. Seahra, and F.~P. Silva, {\it {Cosmological
  perturbations in the DGP braneworld: numeric solution}},  {\em Phys. Rev.}
  {\bf D77} (2008) 083512, [\href{http://xxx.lanl.gov/abs/0711.2563}{{\tt
  arXiv:0711.2563}}].

\bibitem{Ishak:2005zs}
M.~Ishak, A.~Upadhye, and D.~N. Spergel, {\it {Probing cosmic acceleration
  beyond the equation of state: Distinguishing between dark energy and modified
  gravity models}},  {\em Phys. Rev.} {\bf D74} (2006) 043513,
  [\href{http://xxx.lanl.gov/abs/astro-ph/0507184}{{\tt astro-ph/0507184}}].

\bibitem{Thomas:2008tp}
S.~A. Thomas, F.~B. Abdalla, and J.~Weller, {\it {Constraining Modified Gravity
  and Growth with Weak Lensing}},
  \href{http://xxx.lanl.gov/abs/0810.4863}{{\tt arXiv:0810.4863}}.

\bibitem{Dvali:2003rk}
G.~Dvali and M.~S. Turner, {\it {Dark energy as a modification of the Friedmann
  equation}},  \href{http://xxx.lanl.gov/abs/astro-ph/0301510}{{\tt
  astro-ph/0301510}}.

\bibitem{Dvali:2006su}
G.~Dvali, {\it {Predictive Power of Strong Coupling in Theories with Large
  Distance Modified Gravity}},  {\em New J. Phys.} {\bf 8} (2006) 326,
  [\href{http://xxx.lanl.gov/abs/hep-th/0610013}{{\tt hep-th/0610013}}].

\bibitem{Schaefer:2007nf}
B.~M. Schaefer and K.~Koyama, {\it {Spherical collapse in modified gravity with
  the Birkhoff- theorem}},  {\em Mon. Not. Roy. Astron. Soc.} {\bf 385} (2008)
  411--422, [\href{http://xxx.lanl.gov/abs/0711.3129}{{\tt arXiv:0711.3129}}].

\bibitem{Afshordi:2008rd}
N.~Afshordi, G.~Geshnizjani, and J.~Khoury, {\it {Do observations offer
  evidence for cosmological-scale extra dimensions?}},  {\em JCAP} {\bf 0908}
  (2009) 030, [\href{http://xxx.lanl.gov/abs/0812.2244}{{\tt 0812.2244}}].

\bibitem{Kobayashi:2009da}
T.~Kobayashi and H.~Tashiro, {\it {Cluster formation and the Sunyaev-Zel'dovich
  power spectrum in modified gravity: the case of a phenomenologically extended
  DGP model}},  \href{http://xxx.lanl.gov/abs/0903.3738}{{\tt
  arXiv:0903.3738}}.

\bibitem{deRham:2007xp}
C.~de~Rham {\em et.~al.}, {\it {Cascading gravity: Extending the
  Dvali-Gabadadze-Porrati model to higher dimension}},  {\em Phys. Rev. Lett.}
  {\bf 100} (2008) 251603, [\href{http://xxx.lanl.gov/abs/0711.2072}{{\tt
  arXiv:0711.2072}}].

\bibitem{deRham:2007rw}
C.~de~Rham, S.~Hofmann, J.~Khoury, and A.~J. Tolley, {\it {Cascading Gravity
  and Degravitation}},  {\em JCAP} {\bf 0802} (2008) 011,
  [\href{http://xxx.lanl.gov/abs/0712.2821}{{\tt arXiv:0712.2821}}].

\bibitem{deRham:2008qx}
C.~de~Rham, {\it {An Introduction to Cascading Gravity and Degravitation}},
  {\em Can. J. Phys.} {\bf 87} (2009) 201--203,
  [\href{http://xxx.lanl.gov/abs/0810.0269}{{\tt arXiv:0810.0269}}].

\bibitem{Cardone:2009yt}
V.~F. Cardone, A.~Diaferio, and S.~Camera, {\it {Constraining f(R) theories
  with Type Ia Supernovae and Gamma Ray Bursts}},  {\em ArXiv e-prints} (2009)
  [\href{http://xxx.lanl.gov/abs/0907.4689}{{\tt arXiv:0907.4689}}].

\bibitem{Freedman:2000cf}
{\bf HST} Collaboration, W.~L. Freedman {\em et.~al.}, {\it {Final Results from
  the Hubble Space Telescope Key Project to Measure the Hubble Constant}},
  {\em Astrophys. J.} {\bf 553} (2001) 47--72,
  [\href{http://xxx.lanl.gov/abs/astro-ph/0012376}{{\tt astro-ph/0012376}}].

\bibitem{Hicken:2009dk}
M.~Hicken {\em et.~al.}, {\it {Improved Dark Energy Constraints from ~100 New
  CfA Supernova Type Ia Light Curves}},  {\em Astrophys. J.} {\bf 700} (2009)
  1097--1140, [\href{http://xxx.lanl.gov/abs/0901.4804}{{\tt
  arXiv:0901.4804}}].

\bibitem{Cardone:2009mr}
V.~F. Cardone, S.~Capozziello, and M.~G. Dainotti, {\it {An updated Gamma Ray
  Bursts Hubble diagram}},  \href{http://xxx.lanl.gov/abs/0901.3194}{{\tt
  arXiv:0901.3194}}.

\bibitem{Khoury:2009tk}
J.~Khoury and M.~Wyman, {\it {N-Body Simulations of DGP and Degravitation
  Theories}},  {\em Phys. Rev.} {\bf D80} (2009) 064023,
  [\href{http://xxx.lanl.gov/abs/0903.1292}{{\tt arXiv:0903.1292}}].

\bibitem{Smith:2002dz}
{\bf The Virgo Consortium} Collaboration, R.~E. Smith {\em et.~al.}, {\it
  {Stable clustering, the halo model and nonlinear cosmological power
  spectra}},  {\em Mon. Not. Roy. Astron. Soc.} {\bf 341} (2003) 1311,
  [\href{http://xxx.lanl.gov/abs/astro-ph/0207664}{{\tt astro-ph/0207664}}].

\bibitem{Hu:2007pj}
W.~Hu and I.~Sawicki, {\it {A Parameterized Post-Friedmann Framework for
  Modified Gravity}},  {\em Phys. Rev.} {\bf D76} (2007) 104043,
  [\href{http://xxx.lanl.gov/abs/0708.1190}{{\tt arXiv:0708.1190}}].

\bibitem{Hu:2000ee}
W.~Hu, {\it {Weak lensing of the CMB: A harmonic approach}},  {\em Phys. Rev.}
  {\bf D62} (2000) 043007,
  [\href{http://xxx.lanl.gov/abs/astro-ph/0001303}{{\tt astro-ph/0001303}}].

\bibitem{Hu:2001fb}
W.~Hu, {\it {Dark Synergy: Gravitational Lensing and the CMB}},  {\em Phys.
  Rev.} {\bf D65} (2002) 023003,
  [\href{http://xxx.lanl.gov/abs/astro-ph/0108090}{{\tt astro-ph/0108090}}].

\bibitem{Hu:1998az}
W.~Hu and M.~Tegmark, {\it {Weak Lensing: Prospects for Measuring Cosmological
  Parameters}},  {\em Astrophys. J.} {\bf 514} (1999) L65--L68,
  [\href{http://xxx.lanl.gov/abs/astro-ph/9811168}{{\tt astro-ph/9811168}}].

\bibitem{Munshi:2006fn}
D.~Munshi, P.~Valageas, L.~Van~Waerbeke, and A.~Heavens, {\it {Cosmology with
  Weak Lensing Surveys}},  {\em Phys. Rept.} {\bf 462} (2008) 67--121,
  [\href{http://xxx.lanl.gov/abs/astro-ph/0612667}{{\tt astro-ph/0612667}}].

\bibitem{Schmidt:2008hc}
F.~Schmidt, {\it {Weak Lensing Probes of Modified Gravity}},  {\em Phys. Rev.}
  {\bf D78} (2008) 043002, [\href{http://xxx.lanl.gov/abs/0805.4812}{{\tt
  arXiv:0805.4812}}].

\bibitem{Jain:2007yk}
B.~Jain and P.~Zhang, {\it {Observational Tests of Modified Gravity}},  {\em
  Phys.Rev.} {\bf D78} (2008) 063503,
  [\href{http://xxx.lanl.gov/abs/0709.2375}{{\tt arXiv:0709.2375}}].

\bibitem{Tsujikawa:2008in}
S.~Tsujikawa and T.~Tatekawa, {\it {The effect of modified gravity on weak
  lensing}},  {\em Phys.Lett.} {\bf B665} (2008) 325--331,
  [\href{http://xxx.lanl.gov/abs/0804.4343}{{\tt arXiv:0804.4343}}]. * Brief
  entry *.

\bibitem{Camera:2009uz}
S.~Camera, D.~Bertacca, A.~Diaferio, N.~Bartolo, and S.~Matarrese, {\it {Weak
  lensing signal in unified dark matter models}},  {\em Mon. Not. Roy. Astron.
  Soc.} {\bf 399} (2009) 1995--2003,
  [\href{http://xxx.lanl.gov/abs/0902.4204}{{\tt arXiv:0902.4204}}].

\bibitem{Shapiro:2010si}
C.~Shapiro, S.~Dodelson, B.~Hoyle, L.~Samushia, and B.~Flaugher, {\it {Will
  Multiple Probes of Dark Energy find Modified Gravity?}},  {\em Phys.Rev.}
  {\bf D82} (2010) 043520, [\href{http://xxx.lanl.gov/abs/1004.4810}{{\tt
  arXiv:1004.4810}}].

\bibitem{Camera:2010wm}
S.~Camera, T.~D. Kitching, A.~F. Heavens, D.~Bertacca, and A.~Diaferio, {\it
  {Measuring Unified Dark Matter with 3D cosmic shear}},
  \href{http://xxx.lanl.gov/abs/1002.4740}{{\tt arXiv:1002.4740}}.

\bibitem{Martinelli:2010wn}
M.~Martinelli, E.~Calabrese, F.~De~Bernardis, A.~Melchiorri, L.~Pagano, {\em
  et.~al.}, {\it {Constraining Modified Gravity with Euclid}},
  \href{http://xxx.lanl.gov/abs/1010.5755}{{\tt arXiv:1010.5755}}. * Temporary
  entry *.

\bibitem{Abbott:2005bi}
{\bf Dark Energy Survey} Collaboration, T.~Abbott {\em et.~al.}, {\it {The dark
  energy survey}},  \href{http://xxx.lanl.gov/abs/astro-ph/0510346}{{\tt
  astro-ph/0510346}}.

\bibitem{2010AAS...21547704K}
K.~{Kuehn} and {Dark Energy Survey Collaboration}, {\it {Observing Tools for
  the Dark Energy Survey}},  in {\em Bulletin of the American Astronomical
  Society}, vol.~41 of {\em Bulletin of the American Astronomical Society},
  pp.~564--+, Jan., 2010.

\bibitem{2002SPIE.4836..154K}
N.~Kaiser {\em et.~al.}, {\it {Pan-STARRS: A Large Synoptic Survey Telescope
  Array}},  in {\em Society of Photo-Optical Instrumentation Engineers (SPIE)
  Conference Series} ({J.~A.~Tyson and S.~Wolff}, ed.), vol.~4836 of {\em
  Society of Photo-Optical Instrumentation Engineers (SPIE) Conference Series},
  pp.~154--164, Dec., 2002.

\bibitem{2002AAS...20112207K}
N.~{Kaiser} and {Pan-STARRS Team}, {\it {The Pan-STARRS Optical Survey
  Telescope Project}},  vol.~34 of {\em Bulletin of the American Astronomical
  Society}, p.~1304, Dec., 2002.

\bibitem{2008SPIE.7010E..38R}
A.~{Refregier} and M.~{Douspis}, {\it {Summary of the DUNE mission concept}},
  in {\em Society of Photo-Optical Instrumentation Engineers (SPIE) Conference
  Series}, vol.~7010 of {\em Presented at the Society of Photo-Optical
  Instrumentation Engineers (SPIE) Conference}, Aug., 2008.

\bibitem{Refregier:2010ss}
A.~Refregier {\em et.~al.}, {\it {Euclid Imaging Consortium Science Book}},
  {\em arXiv:1001.0061} (2010) [\href{http://xxx.lanl.gov/abs/1001.0061}{{\tt
  arXiv:1001.0061}}].

\bibitem{Deffayet:2000uy}
C.~Deffayet, {\it {Cosmology on a brane in Minkowski bulk}},  {\em Phys. Lett.}
  {\bf B502} (2001) 199--208,
  [\href{http://xxx.lanl.gov/abs/hep-th/0010186}{{\tt hep-th/0010186}}].

\bibitem{Schaefer:2006pa}
B.~E. Schaefer, {\it {The Hubble Diagram to Redshift >6 from 69 Gamma-Ray
  Bursts}},  {\em Astrophys. J.} {\bf 660} (2007) 16--46,
  [\href{http://xxx.lanl.gov/abs/astro-ph/0612285}{{\tt astro-ph/0612285}}].

\bibitem{Eisenstein:2005su}
{\bf SDSS} Collaboration, D.~J. Eisenstein {\em et.~al.}, {\it {Detection of
  the Baryon Acoustic Peak in the Large-Scale Correlation Function of SDSS
  Luminous Red Galaxies}},  {\em Astrophys. J.} {\bf 633} (2005) 560--574,
  [\href{http://xxx.lanl.gov/abs/astro-ph/0501171}{{\tt astro-ph/0501171}}].

\bibitem{Tegmark:2006az}
{\bf SDSS} Collaboration, M.~Tegmark {\em et.~al.}, {\it {Cosmological
  Constraints from the SDSS Luminous Red Galaxies}},  {\em Phys. Rev.} {\bf
  D74} (2006) 123507, [\href{http://xxx.lanl.gov/abs/astro-ph/0608632}{{\tt
  astro-ph/0608632}}].

\bibitem{Lewis:1999bs}
A.~Lewis, A.~Challinor, and A.~Lasenby, {\it {Efficient Computation of CMB
  anisotropies in closed FRW models}},  {\em Astrophys. J.} {\bf 538} (2000)
  473--476, [\href{http://xxx.lanl.gov/abs/astro-ph/9911177}{{\tt
  astro-ph/9911177}}].

\bibitem{Fang:2008sn}
W.~Fang, W.~Hu, and A.~Lewis, {\it {Crossing the Phantom Divide with
  Parameterized Post- Friedmann Dark Energy}},  {\em Phys. Rev.} {\bf D78}
  (2008) 087303, [\href{http://xxx.lanl.gov/abs/0808.3125}{{\tt
  arXiv:0808.3125}}].

\bibitem{Fang:2008kc}
W.~Fang {\em et.~al.}, {\it {Challenges to the DGP Model from Horizon-Scale
  Growth and Geometry}},  {\em Phys. Rev.} {\bf D78} (2008) 103509,
  [\href{http://xxx.lanl.gov/abs/0808.2208}{{\tt arXiv:0808.2208}}].

\bibitem{Lue:2003ky}
A.~Lue, R.~Scoccimarro, and G.~Starkman, {\it {Differentiating between Modified
  Gravity and Dark Energy}},  {\em Phys. Rev.} {\bf D69} (2004) 044005,
  [\href{http://xxx.lanl.gov/abs/astro-ph/0307034}{{\tt astro-ph/0307034}}].

\bibitem{Koyama:2006ef}
K.~Koyama, {\it {Structure formation in modified gravity models alternative to
  dark energy}},  {\em JCAP} {\bf 0603} (2006) 017,
  [\href{http://xxx.lanl.gov/abs/astro-ph/0601220}{{\tt astro-ph/0601220}}].

\bibitem{Eisenstein:1997jh}
D.~J. Eisenstein and W.~Hu, {\it {Power Spectra for Cold Dark Matter and its
  Variants}},  {\em Astrophys. J.} {\bf 511} (1999) 5,
  [\href{http://xxx.lanl.gov/abs/astro-ph/9710252}{{\tt astro-ph/9710252}}].

\bibitem{Koyama:2009me}
K.~Koyama, A.~Taruya, and T.~Hiramatsu, {\it {Non-linear Evolution of Matter
  Power Spectrum in Modified Theory of Gravity}},  {\em Phys. Rev.} {\bf D79}
  (2009) 123512, [\href{http://xxx.lanl.gov/abs/0902.0618}{{\tt
  arXiv:0902.0618}}].

\bibitem{Oyaizu:2008tb}
H.~Oyaizu, M.~Lima, and W.~Hu, {\it {Non-linear evolution of f(R) cosmologies
  II: power spectrum}},  {\em Phys. Rev.} {\bf D78} (2008) 123524,
  [\href{http://xxx.lanl.gov/abs/0807.2462}{{\tt arXiv:0807.2462}}].

\bibitem{Schmidt:2009sg}
F.~Schmidt, {\it {Self-Consistent Cosmological Simulations of DGP Braneworld
  Gravity}},  {\em Phys. Rev.} {\bf D80} (2009) 043001,
  [\href{http://xxx.lanl.gov/abs/0905.0858}{{\tt arXiv:0905.0858}}].

\bibitem{Schmidt:2009yj}
F.~Schmidt, W.~Hu, and M.~Lima, {\it {Spherical Collapse and the Halo Model in
  Braneworld Gravity}},  {\em Phys. Rev.} {\bf D81} (2010) 063005,
  [\href{http://xxx.lanl.gov/abs/0911.5178}{{\tt arXiv:0911.5178}}].

\bibitem{Schmidt:2009sv}
F.~Schmidt, {\it {Cosmological Simulations of Normal-Branch Braneworld
  Gravity}},  {\em Phys. Rev.} {\bf D80} (2009) 123003,
  [\href{http://xxx.lanl.gov/abs/0910.0235}{{\tt arXiv:0910.0235}}].

\bibitem{Bartelmann:1999yn}
M.~Bartelmann and P.~Schneider, {\it {Weak Gravitational Lensing}},  {\em Phys.
  Rept.} {\bf 340} (2001) 291--472,
  [\href{http://xxx.lanl.gov/abs/astro-ph/9912508}{{\tt astro-ph/9912508}}].

\bibitem{Kaiser:1996tp}
N.~Kaiser, {\it {Weak Lensing and Cosmology}},  {\em Astrophys. J.} {\bf 498}
  (1998) 26, [\href{http://xxx.lanl.gov/abs/astro-ph/9610120}{{\tt
  astro-ph/9610120}}].

\bibitem{Kaiser:1991qi}
N.~Kaiser, {\it {Weak gravitational lensing of distant galaxies}},  {\em
  Astrophys. J.} {\bf 388} (1992) 272.

\bibitem{Fu:2007qq}
L.~Fu {\em et.~al.}, {\it {Very weak lensing in the CFHTLS Wide: Cosmology from
  cosmic shear in the linear regime}},  {\em Astron. Astrophys.} {\bf 479}
  (2008) 9--25, [\href{http://xxx.lanl.gov/abs/0712.0884}{{\tt
  arXiv:0712.0884}}].

\bibitem{Blake:2004tr}
C.~Blake and S.~Bridle, {\it {Cosmology with photometric redshift surveys}},
  {\em Mon. Not. Roy. Astron. Soc.} {\bf 363} (2005) 1329--1348,
  [\href{http://xxx.lanl.gov/abs/astro-ph/0411713}{{\tt astro-ph/0411713}}].

\bibitem{Hawken:2009fr}
A.~J. Hawken and S.~L. Bridle, {\it {Gravitational flexion by elliptical dark
  matter haloes}},  {\em Mon. Not. Roy. Astron. Soc.} {\bf 400} (2009)
  1132--1138, [\href{http://xxx.lanl.gov/abs/0903.3938}{{\tt
  arXiv:0903.3938}}].

\end{thebibliography}\endgroup

\end{document}